\documentclass[letterpaper]{jpconf}
\usepackage{graphicx}
\RequirePackage{hepparticles}
\RequirePackage{hepnames}
\RequirePackage{hepnicenames}
\usepackage{textcomp}
\usepackage{citesort}
\usepackage{multirow}
\usepackage{rotating}
\usepackage{subfigure}
\newcommand{\lhcb}{LHCb}
\newcommand{\dalitz}{Dalitz}
\newcommand{\mygamma}{$\gamma$}
\newcommand{\myrb}{$\text{r}_{\PBpm}$}
\newcommand{\myrbz}{$\text{r}_{\PBzero}$}
\newcommand{\mydeltab}{$\delta_{\PBpm}$}
\newcommand{\mydeltabz}{$\delta_{\PBzero}$}

\newcommand{\decayBtoDK}{\mbox{\HepProcess{\PB\HepTo\PD\PK}}}

\newcommand{\decayBtoDhhK}{\mbox{\HepProcess{\PBpm^{(0)}\HepTo\PD(hh)\PKpm^{(*0)}}}}

\newcommand{\decayBtoDkspipiK}{\mbox{\HepProcess{\PBpm\HepTo\PD(\PKs\Ppiplus\Ppiminus)\PKpm}}}

\newcommand{\decayBstoDsK}{\mbox{\HepProcess{\PBs\HepTo\PDs\PK}}}

\newcommand{\decayBztoDpi}{\mbox{\HepProcess{\PBzero\HepTo\PD\Ppi}}}

\newcommand{\decayBztopipi}{\mbox{\HepProcess{\PBzero\HepTo\Ppi\Ppi}}}

\newcommand{\decayBstoKK}{\mbox{\HepProcess{\PBs\HepTo\PK\PK}}}

\newcommand{\decayButoKpipi}{\mbox{\HepProcess{\PBplus\HepTo\Ppiplus\Ppiminus}}}

\newcommand{\decayBztoKspipi}{\mbox{\HepProcess{\PBzero\HepTo\PKs\Ppiplus\Ppiminus}}}

\begin{document}
\title{Prospects for \mygamma\ measurements at \lhcb}

\author{Ying Ying Li (On behalf of the \lhcb\ collaboration)}

\address{University of Cambridge, United Kingdom}

\begin{abstract}
\lhcb\ is the dedicated B physics experiment at the LHC and is due to start data taking later this year. Its goal is to search for new physics in very rare processes and make precision measurements of CP violation in \PB decays. The CKM angle \mygamma\ plays an important role in flavour physics in the Standard Model. \lhcb\ will exploit the large variety of \PB hadrons produced by the 14 TeV pp collisions, performing \mygamma\ measurements to the precision of a few degrees. Here, we will present a summary of the expected \mygamma\ sensitivities \lhcb\ will reach during its first years of data taking, with contributions from several strategies in both tree and loop processes.
\end{abstract}

\section{Introduction}
\subsection{The LHCb experiment}
\lhcb\ is the LHC beauty experiment~\cite{lhcbdet08,Voss08}. It is a single-arm forward spectrometer taking advantage of the forward \HepProcess{\Pbeauty\APbeauty} productions at the 14 TeV collisions. \lhcb\ will see an integrated \HepProcess{\Pbeauty\APbeauty} cross-section of 500 $\mu$b, equating to $\sim 10^{12}$ \HepProcess{\Pbeauty\APbeauty} pairs per year and producing a full range of \Pbeauty-hadron species. The sub-detectors include an excellent tracking system, which will give life-time resolutions of $\sim 200$ $\mu$m and a \PB mass resolution of $\sim 14$ MeV. The \lhcb\ RICH system for hadronic particle identification (PID) will give a kaon PID efficiency of $\sim 96$\% and a pion mis-identification of $\sim 5\%$.

\subsection{CKM angle \mygamma}
The source of CP-violation in the Standard Model (SM) arises from an irreducible complex phase of the CKM matrix, which can be represented by the unitarity triangle (UT) with angles of $\alpha$, $\beta$ and \mygamma. The current experimental constraints placed on the least well know angle \mygamma\ is
\begin{equation}
    \gamma = (70^{+27}_{-30})\text{\textdegree} \quad \text{and} \quad (67.8^{+4.2}_{-3.9})\text{\textdegree}, \nonumber
\end{equation}
from direct (tree diagram processes) and indirect \mygamma\ measurements respectively~\cite{ckmfitter08}, showing excellent agreement with SM predictions. The indirect measurement from the rest of the UT measurements is dominated by the measurement of the mixing side in the UT with contributions from loop processes. \lhcb\ aims to measure all three angles and the apex of the UT independently, testing the SM to its limits and probing new physics contributions in loop processes. The angle \mygamma\ can be extracted from both loop and tree processes. However, in order to  determine new physics contributions in loops, we require a precision \mygamma\ measurement of only a few degrees from tree processes. The following sections summarises the current \lhcb\ sensitivity estimates for the measurement of \mygamma\ from both tree and loop diagrams.
 
\section{$\gamma$ from tree processes}

\subsection{\decayBtoDK\ strategies}
The theoretically cleanest and powerful way of measuring \mygamma\ from trees is through the \decayBtoDK\ decays. For \PBpm and \PBzero we only have contributions from two interfering trees, \HepProcess{\Pbeauty\HepTo\Pcharm\PW} and \HepProcess{\Pbeauty\HepTo\Pup\PW} ($\propto e^{i\gamma}$). In the case of \PBpm, the two tree diagrams are colour and CKM favoured and suppressed contributions, while for the \PBzero case both trees are colour suppressed. Three parameters of the interference between the two trees can be extracted if both \PDzero and \APDzero decay to the same final state, \myrb\ (\myrbz), the ratio between the two trees, \mygamma, the CP-violating weak phase and \mydeltab\ (\mydeltabz), the CP-conserving strong phase difference between \PBplus (\PBzero) and \PBminus (\APBzero). The methods of extraction of the three parameters is determined by the type of \PD (\HepProcess{\PDzero/\APDzero}) decay.

\subsubsection{\decayBtoDhhK\ decays with ADS/GLW method}
In the two-body \HepProcess{\PD\HepTo hh} decay, the so-called ADS method can be used in the case when \PD decays to doubly Cabibbo favoured and suppress states of \HepProcess{\PKpm\Ppimp}~\cite{ADS97,ADS01}, or the GLW method can be used in the case of CP eigenstates, \HepProcess{\PKplus\PKminus} and \HepProcess{\Ppiplus\Ppiminus}~\cite{GLW91,GLW1991}. Combining the ADS and GLW modes results in a total of six rates consisting of three \PB parameters \mygamma, $\text{r}_{\PB^{\pm(0)}}$, $\delta_{\PB^{\pm(0)}}$ and two additional \PD parameters $\text{r}_{\PK\Ppi}$, the ratios between the favoured and suppressed \PD decay tree diagrams and $\delta_{\PK\Ppi}$ the \PD strong phase. These additional \PD parameters can be constrained from external measurements made at \mbox{CLEO-c}~\cite{rosner2008,asner2008}. Current studies estimate a \lhcb\ \mygamma\ sensitivity with an integrated luminosity of 2 $\text{fb}^{-1}$ (a full year of data taking) of \HepProcess{\PBpm\HepTo\PD(hh)\PKpm}: $\sigma_{\gamma} = (10.8-13.8)\text{\textdegree}$ for $\delta_{\PK\Ppi} = -(130-190)\text{\textdegree}$ and \HepProcess{\PBzero\HepTo\PD(hh)\PK^{*0}}: $\sigma_{\gamma} =  (5.2-12.7)\text{\textdegree}$ for $\delta_{\PBzero} = (0-180)\text{\textdegree}$~\cite{Libby08}. The addition of the ADS \HepProcess{\PBpm\HepTo\PD(\PK\Ppi\Ppi\Ppi)\PKpm} decay mode can improve the above \mygamma\ sensitivity to (7-10)\textdegree with 2 $\text{fb}^{-1}$ of data~\cite{Libby08}.

\subsubsection{\decayBtoDkspipiK\ decays with GGSZ (\dalitz) method}
\begin{figure}[h]
\centering
\begin{sideways}
\includegraphics[width=14pc]{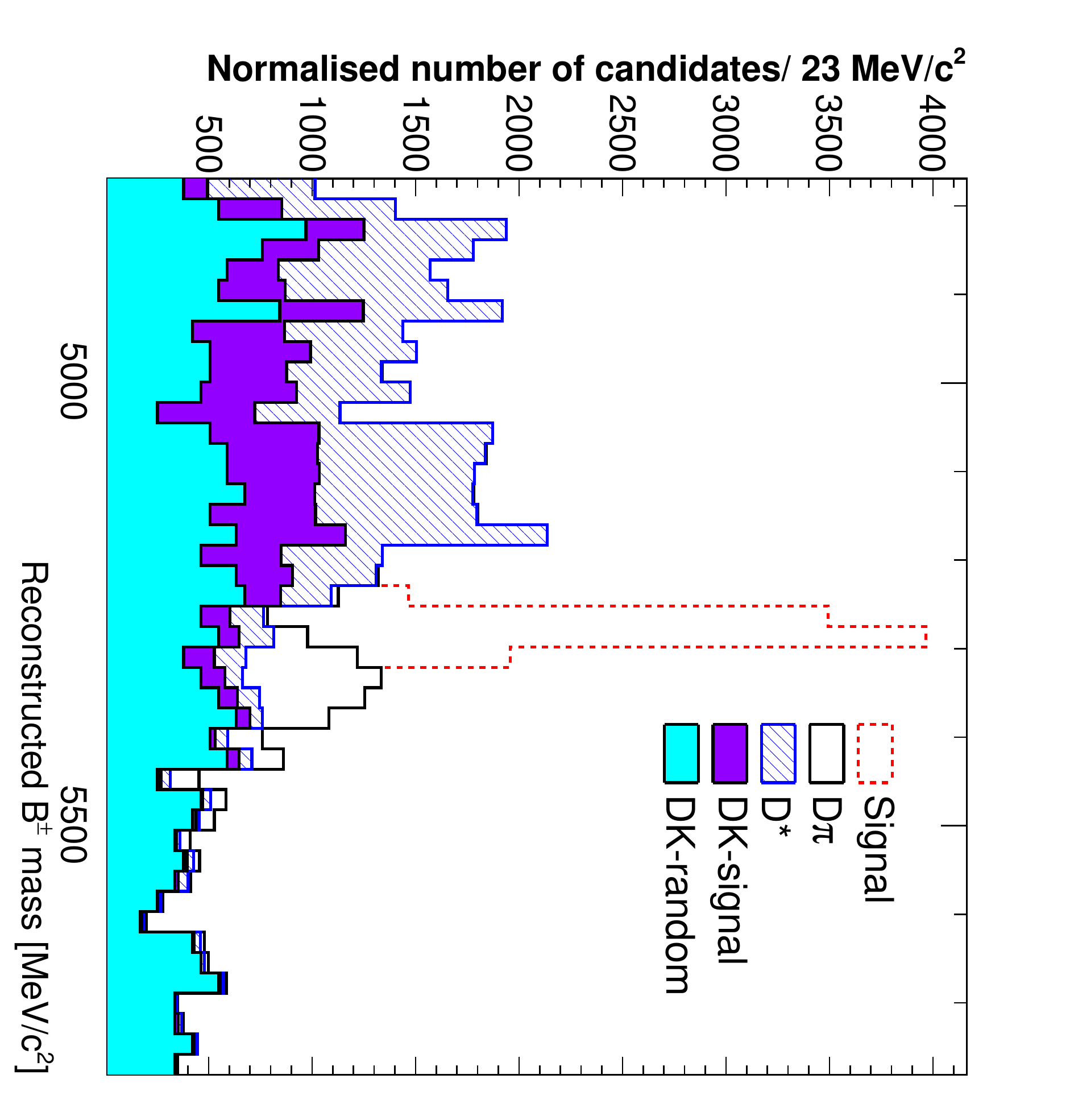}\hspace{1pc}%
\end{sideways}
\begin{minipage}[b]{14pc}\caption{\label{label}Signal and background \PB mass distributions for the \decayBtoDkspipiK\ decay, where $\PD\PK$-random and $\PD\Ppi$ backgrounds are discussed in the text, the \PDst background are events where a real \PD from a \PDst is paired with a fake kaon from the same event, the $\PD\PK$-signal is where a real \PD that is reconstructed with a fake bachelor kaon from the same event.}
\end{minipage}
\label{fig:bkgs}
\end{figure}

\begin{figure}[h]
\centering
\subfigure[]{
\begin{sideways}
\includegraphics[width=14pc]{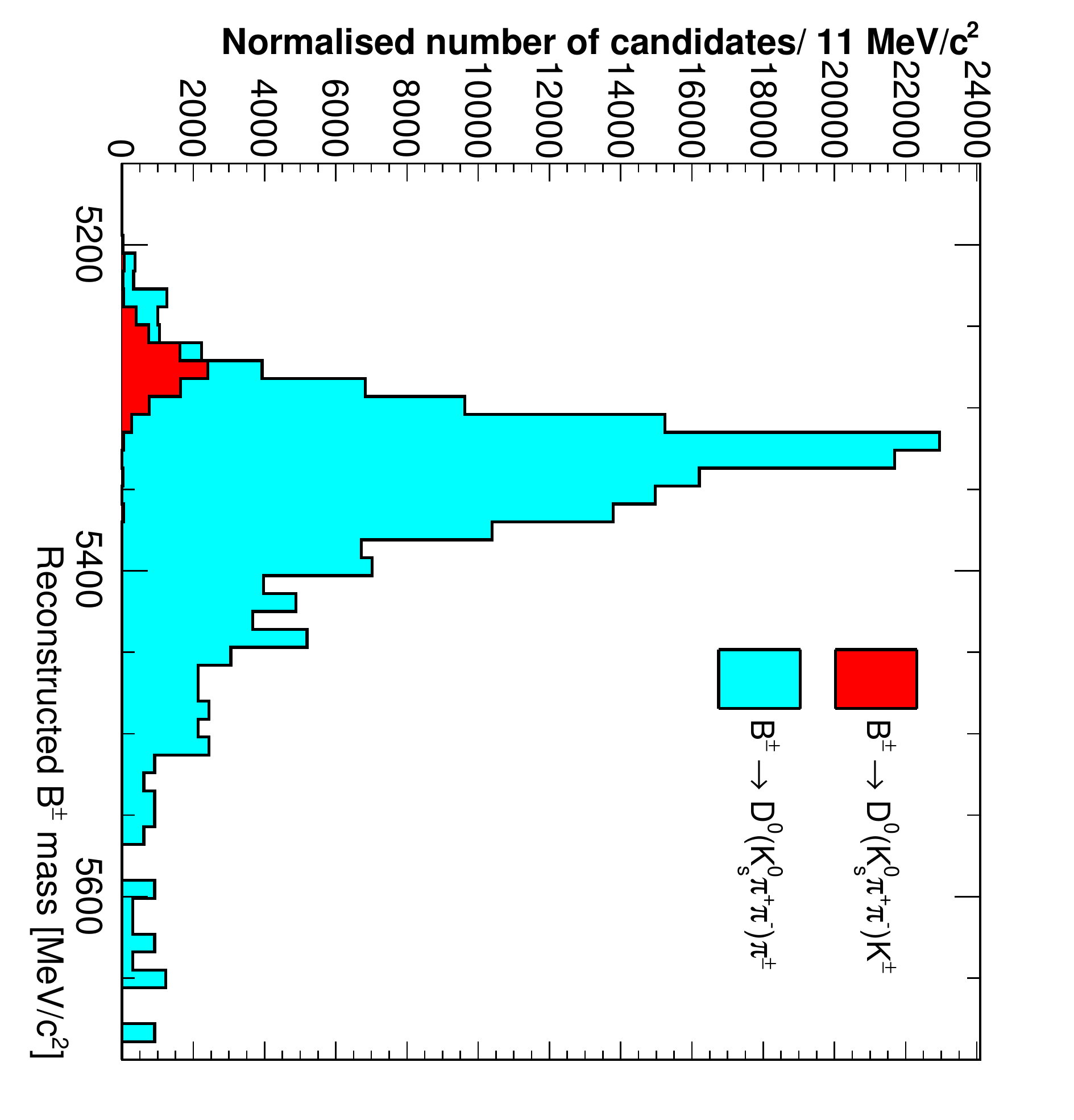}\hspace{1pc}%
\label{fig:nopid}
\end{sideways}
}
\subfigure[]{
\begin{sideways}
\includegraphics[width=14pc]{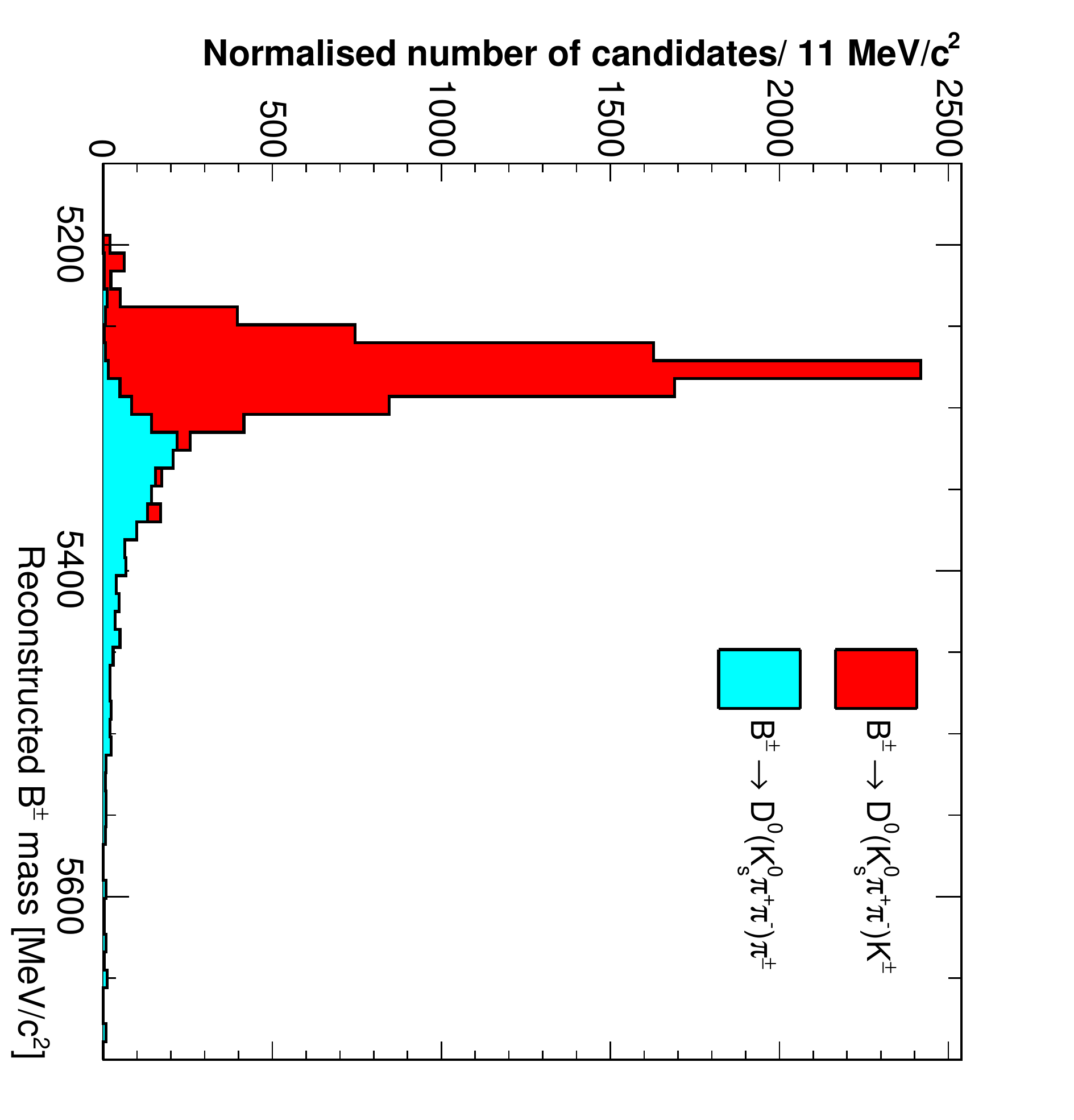}\hspace{1pc}%
\label{fig:withpid}
\end{sideways}
}
\begin{minipage}[b]{6pc}\caption{\label{label}Reconstructed \PB mass in the \HepProcess{\PBpm\HepTo\PD\PKpm} hypothesis for \HepProcess{\PBpm\HepTo\PD\PKpm} and \HepProcess{\PBpm\HepTo\PD\Ppipm} events, without RICH PID \Fref{fig:nopid} and with PID \Fref{fig:withpid}.}
\end{minipage}
\label{fig:PID}
\end{figure}

The GGSZ (\dalitz) method can be used in the case of three-body \HepProcess{\PD\HepTo\PKs\Ppiplus\Ppiminus} decay~\cite{GGSZ03}, where the \mygamma\ sensitivity can be seen as a difference in the density of events over the resonance rich \dalitz\ plane of the \PD decay. The extraction of \mygamma\ from the resonances can be made either with a unbinned isobar model approach, as used by the \PB-factories~\cite{belle,babar}, or a binned model independent method, where the bins are determined by the \PD strong phase, $\delta_{\PD}$. At the 2 $\text{fb}^{-1}$ level \lhcb\ can expect an annual signal yield of $\sim 5,000$ events, with a dominant background estimated to be the combinatoric background of a real \PD reconstructed with a fake kaon from the underlying event ($\PD\PK$-random), at a background to signal ratio of $B/S = 0.35\pm 0.03$~\cite{Gibson08}, as shown in \Fref{fig:bkgs}. The second dominant background, \HepProcess{\PBpm\HepTo\PD(\PKs\Ppiplus\Ppiminus)\Ppipm} ($\PD\Ppi$ background), where the pion is misidentified as a kaon is greatly reduced by the RICH PID information, to a $B/S < 0.095$ at the 90\%\ confidence level, see \Fref{fig:PID}. The expected \mygamma\ sensitivity with 2 $\text{fb}^{-1}$ of data and $\text{\myrb} = 0.1$ using the Belle and Babar Isobar model: $\sigma_{\gamma} = 9.8\text{\textdegree}$ with a $\sim 7$\textdegree\ model error~\cite{Gibson07} and binned model independent: $\sigma_{\gamma} = 12.8\text{\textdegree}$ with a $\sim$(1-2)\textdegree\ CLEO-c statistical error~\cite{Libby07}, where some sensitivity is lost in the binning, making the model dependent method the most sensitive for first data, while for $> 2$ $\text{fb}^{-1}$, the binned model independent method will become the more sensitive method~\cite{Thomas08}.

\subsection{Time-dependent \decayBstoDsK\ and \decayBztoDpi\ decays}
A time dependent \mygamma\ and $\phi_{\Pstrange}$ (\PBs mixing angle) extraction can be made from the flavour tagged \decayBstoDsK\ decay, where CP-violation arises from the mixing and the high amplitude interference between the \HepProcess{\Pbeauty\HepTo\Pcharm\PW} and \HepProcess{\Pbeauty\HepTo\Pup\PW} trees. Using $\text{r}_{\PBs} \sim 0.4$ and assuming that $\phi_{\Pstrange}$ will be constrained by other \lhcb\ measurements (e.g. \HepProcess{\PBs\HepTo\PJpsi\Pphi}) gives a signal annual yield of $\sim$6,200 events, with a $B/S = 0.7$~\cite{borel07} and \mygamma\ sensitivity of $\sigma_{\gamma} \approx 10.3\text{\textdegree} (4.6\text{\textdegree})$ with 2 (10) $\text{fb}^{-1}$ of data~\cite{cohen07}.

\section{\mygamma\ from loop processes}

\subsection{\decayBztopipi\ and \decayBstoKK\ decays}
In the case of the \decayBztopipi\ and \decayBstoKK\ decays, \mygamma\ sensitivity arises from the \HepProcess{\Pbeauty\HepTo\Pup\PW} tree diagrams, however the presence of loop diagrams in the interference can perturb the result. In these decays a fit is made to the time-dependent CP-asymmetry with four U-spin asymmetry observables, a direct and a mixing asymmetry from each decay mode~\cite{Fleischer1990}. Including a 20\%\ U-spin asymmetry gives a \mygamma\ sensitivity of $\sigma_{\gamma} = 10\text{\textdegree} (5\text{\textdegree})$ with 2 (10) $\text{fb}^{-1}$ of data.

\subsection{\decayButoKpipi\ and \decayBztoKspipi\ decays}
A first study of the three-body decays of \decayButoKpipi\ and \decayBztoKspipi\ has also been performed, where there are contributions from \HepProcess{\Pbeauty\HepTo\Pup\PW} tree diagram in the \PBzero decay, and two loop diagrams, from \PBplus and \PBzero, \HepProcess{\Pbeauty\HepTo\Pstrange\PW}. The sensitivity to \mygamma\ can be extracted via a two step \dalitz\ analysis, where the penguin contributions from the \decayButoKpipi\ decay is first deduced from a \dalitz\ anisotropy of the CP-asymmetries and phase differences. This is followed by subtracting the penguin contribution, of the first step, from the \dalitz\ analysis of \decayBztoKspipi\ to obtain the tree contribution and therefore a sensitivity to \mygamma. The expected annual yield for the \PBplus and \PBzero decays are $\sim$494,000 and $\sim$90,000 events respectively, giving a \mygamma\ sensitivity of $\sigma_{\gamma} \sim 5\text{\textdegree}$ with 2 $\text{fb}^{-1}$ of data~\cite{Guerrer08}.

\section{Conclusions}
\lhcb\ has a rich \mygamma\ measurement program, measuring \mygamma\ from trees and loops with an aim to combine and disentangle new physics from the SM predictions. Measurements from tree diagrams to a few degrees will provide a \textquotedblleft standard candle\textquotedblright\ \mygamma\ measurement, with results from CLEO-c playing an important role in determining the parameters of the underlying \PD decay. Combining the studies so far from trees diagrams in a global fit show \lhcb\ can expect a SM baseline \mygamma\ sensitivity of $\sigma_{\gamma} \sim (4-5)\text{\textdegree}$ $(2-3)$\textdegree\ for 2 (10) $\text{fb}^{-1}$ of data, with the sensitivity contribution from each channel dependent on the \PBzero strong phase, $\delta_{\PBzero}$, as shown in \Tref{tab:senscontrib}. The estimations from loops processes gives \mbox{$\sigma_{\gamma} \sim 10\text{\textdegree}$ $(5\text{\textdegree})$} at 2 (10) $\text{fb}^{-1}$, with promising first results of $\sigma_{\gamma}\sim 5 \text{\textdegree}$ from the \HepProcess{\PB\HepTo hhh} decays at 2 $\text{fb}^{-1}$.
\begin{table}[h]
\caption{Contribution to global \mygamma\ sensitivity from various tree diagram decay modes~\cite{Libby08}.}
\centering
\begin{tabular}{lcc}
\br
Decay &$\delta_{\PBzero} = 0\text{\textdegree}$ &$\delta_{\PBzero} = 45\text{\textdegree}$\\
\mr
\HepProcess{\PBpm\HepTo\PD(hh)\PKpm}, &\multirow{2}*{25\%}  &\multirow{2}*{38\%}\\
\HepProcess{\PBpm\HepTo\PD(\PK 3\Ppi)\PKpm} \\
\decayBtoDkspipiK &12\% &25\% \\
\HepProcess{\PBzero\HepTo\PD(hh)\PK^{*0}} &44\% &8\% \\
\decayBstoDsK &3\% &5\% \\
\br
\end{tabular}
\label{tab:senscontrib}
\end{table}

\section*{Acknowledgements}
The author would like to thank the organisers of LLWI for this opportunity and the \lhcb\ collaborators for their invaluable suggestions in the preparation of these proceedings.

\section*{References}
\bibliographystyle{iopart-num}
\bibliography{iopart-num}
\end{document}